\documentclass[twocolumn,showpacs,preprintnumbers,amsmath,amssymb]{revtex4}

\usepackage{graphicx}
\usepackage{dcolumn}
\usepackage{bm}

\begin{document}

\preprint{APS/123-QED}

\title{Effect of Elastic Deformations on the Critical Behavior
of Three-Dimensional Systems with Long-Range Interaction}

\author{S.V. Belim}
 \email{belim@univer.omsk.su}
\affiliation{%
Omsk State University, 55-a, pr. Mira, Omsk, Russia, 644077
\textbackslash\textbackslash
}%

\date{\today}

\begin{abstract}
A field-theoretical description of the behavior of compressible Ising systems with long-range interactions is
presented. The description is performed in the two-loop approximation in three dimensions with the use of the
Padé–Borel resummation technique. The renormalization group equations are analyzed, and the fixed points
that determine the critical behavior of the system are found. It is shown that the effect of elastic deformations
on a system with a long-range interaction causes changes in its critical, as well as multicritical, behavior.

\end{abstract}

\maketitle

In compressible systems, the relation of the order
parameter to elastic deformations plays an important
role. Earlier [1] it was shown that, in the case of an elastically
isotropic body, the critical behavior of a compressible
system with a quadratic striction is unstable
with respect to the relation of the order parameter to
acoustic modes, and a first-order phase transition close
to a second-order one is realized. However, the conclusions
formulated in the cited paper [1] hold only for low
pressures. As was shown later [2], at high pressures
beginning from a certain tricritical point $P_t$, the deformations
induced by the external pressure affect the system
to a greater extent and lead to a change of sign of
the effective interaction constant for the order parameter
fluctuations and, as a consequence, to a change in
the order of the phase transition. In this case, according
to [2], a homogeneous compressible system is characterized
by two types of tricritical behavior with a
fourth-order critical point formed as the point of intersection
of the two tricritical curves. Calculations performed
in terms of the two-loop approximation [3] confirmed the presence of two types of
tricritical behavior for Ising systems and provided values of the tricritical
indices.

In structural phase transitions that occur in the
absence of the piezoelectric effect, in the paraphase the
elastic strains play the role of a secondary order parameter
whose fluctuations are not critical in most cases.

The effect of the long-range interaction described by
the power law $1/r^{-D-\sigma}$ at long distances was studied
analytically in terms of the $\varepsilon$-expansion [4–6] and
numerically by the Monte Carlo method [7–9] in two
and three dimensions. It was found that the long-range
interaction considerably affects the critical behavior of
Ising systems for the parameter values $\sigma<2$. A recent
study carried out for a three-dimensional space in the
two-loop approximation [10] confirmed the prediction
of the $\varepsilon$-expansion for systems with long-range interactions.

This paper describes the critical and tricritical
behavior of three-dimensional compressible systems
by taking into account the effect of the long-range interaction
with different values of the parameter $\sigma$.

For a homogeneous Ising-like model with elastic
deformations and a long-range interaction, the Hamiltonian
can be represented in the form

\begin{eqnarray}
&&H_0=\int d^Dx[\frac{1}{2}(\tau_1+\nabla^\sigma)\vec{S}(x)^2 +
+\frac{u_0}{4!}(\vec{S}(x)^2)^2] \nonumber\\
&&+\int d^Dx[a_1(\sum\limits_{\alpha
=1}^3u_{\alpha \alpha }(x))^2+a_2\sum\limits_{\alpha ,\beta
=1}^3u_{\alpha \beta }^2]+\nonumber\\
 &+&\frac 12a_3\int d^Dx\vec{S}(x)^2 (\sum\limits_{\alpha =1}^3u_{\alpha\alpha }(x))
\end{eqnarray}
where $S(x)$ is the scalar order parameter, $u_0$ is a positive
constant, $\tau_0\sim|T-T_c|$, $T_c$ is the phase transition temperature,
$u_{\alpha \beta}$ is the strain tensor, $a_1$ and $a_2$ are the elastic
constants of the crystal, and $a_3$ is the quadratic striction
parameter. Let us change to the Fourier transforms of
the variables in Eq. (1) and perform integration with
respect to the components depending on the nonfluctuating
variables, which do not interact with the order parameter $S(x)$
Then, introducing, for convenience, the new variable
$y(x)=\sum\limits_{\alpha=1}^3u_{\alpha\alpha}(x)$, we obtain the
Hamiltonian of the system in the form
\begin{eqnarray}
&&H_0=\frac 12\int d^Dq(\tau _0+q^\sigma)S_qS_{-q}\nonumber\\
&&+ \frac{u_0}{4!}\int d^D{q_i}S_{q1}S_{q2}S_{q3}S_{-q1-q2-q3}+\nonumber\\
&&+a_3\int d^Dqy_{q1}S_{q2}S_{-q1-q2}
+\frac{a_3^{(0)}}\Omega y_0\int d^DqS_qS_{-q}\nonumber\\
&&+\frac 12a_1 \int d^Dqy_qy_{-q} +\frac
12\frac{a_1^{(0)}}\Omega y_0^2
\end{eqnarray}
In Eq. (2), the components $y_0$ describing the homogeneous
strains are separated. According to [1], such a
separation is necessary, because the inhomogeneous
strains $y_q$ are responsible for the acoustic phonon
exchange and lead to long-range interactions, which are
absent in the case of homogeneous strains.
For the system under study, let us determine the
effective Hamiltonian that depends only on the strongly
fluctuating order parameter S
\begin{eqnarray}\label{usr}
\exp \{-H[S]\}=B\int \exp \{-H_{R}[S,y]\}\prod dy_q
\end{eqnarray}
If the experiment is performed at constant volume, $y_0$
is a constant and the integration in Eq. (3) is performed
with respect to only the inhomogeneous strains, while
the homogeneous strains do not contribute to the effective
Hamiltonian. If the experiment occurs at constant
pressure, the term $P\Omega$ is added to the Hamiltonian, the
volume is represented in terms of the strain tensor components
as
\begin{eqnarray}\label{vol}
\Omega=\Omega_0 [1+\sum\limits_{\alpha =1}u_{\alpha\alpha}+
\sum\limits_{\alpha \neq
\beta}u_{\alpha\alpha}u_{\beta\beta}+O(u^3)]
\end{eqnarray}
and the integration in Eq. (3) is performed with respect
to the homogeneous strains as well. According to [11],
the inclusion of quadratic terms in Eq. (4) may be
important when dealing with high pressures and crystals
with large striction effects. The neglect of these
quadratic terms restricts the applicability of the results
obtained by Larkin and Pikin [1] to the case of low
pressures. Thus, the Hamiltonian has the form
\begin{eqnarray}\label{ogam}
&&H=\frac 12\int d^Dq(\tau _0+q^\sigma)S_qS_{-q}\nonumber\\
&&+(u_0- \frac{z_0}{2})\int d^Dq_1d^Dq_2d^Dq_3S_{q1}S_{q2}S^a_{q3}S_{-q1-q2-q3}\nonumber\\
&&+\frac{1}{2\Omega}(z_0 - w_0)\int d^Dq_1d^Dq_2S_{q1}S_{-q1}S_{q2}S_{-q2},\\
&&z_0 =a_1^2/(4a_3),  \ \  w_0 =a_1^{(0)2}/(4a_3^{(0)}),  \nonumber
\end{eqnarray}
The effective interaction parameter $v_0=u_0-z_0/2$
that appears in the Hamiltonian due to striction effects,
which are determined by the parameter $z_0$, can take not
only positive but also negative values. As a result, the
Hamiltonian describes both first-order and secondorder
phase transitions. At $v_0= 0$, the system exhibits a
tricritical behavior. In its turn, the effective interaction
determined in Hamiltonian (5) by the parameter difference
$z_0–w_0$ may cause a second-order phase transition
in the system when $z_0–w_0> 0$ and a first-order phase
transition when $z_0–w_0< 0$. This form of the effective
Hamiltonian suggests that a higher order critical point
can be realized in the system as the point of intersection
of the tricritical curves when the conditions $v_0= 0$ and
$z_0=w_0$ are simultaneously satisfied [2]. It should be
noted that, with the tricritical condition $z_0=w_0$, Hamiltonian
(5) of the model under consideration is isomorphic
with the Hamiltonian of a rigid homogeneous system.

In the framework of the field-theoretical approach [12], the
asymptotic critical behavior and the structure of the phase
diagrams in the fluctuation region are determined by the
Callan–Symanzik renormalization group equation for the vertex
parts of the irreducible Green functions. To calculate the
$\beta-$ and $\gamma$-functions as the functions involved in the
Callan–Symanzik equation for renormalized interaction vertices
$u$, $a_1$, and $a_1^(0)$, or complex vertices $z= a_1^2/4a_3$,
$w=a_1^{(0)2} /4a_3^{(0)}$, and $v = u – 12z$, which are more
convenient for the determination of critical and tricritical
behavior, a standard method based on the Feynman diagram technique
and on the renormalization procedure [13] was used with the
propagator $G(\vec{k})=1/{\tau+|\vec{k}|^\sigma}$. As a result, the following
expressions were obtained for the $\beta-$ and $\gamma$- functions in the
two-loop approximation:
\begin{eqnarray}
    \beta_v&=&-(2\sigma-D)v\Big[1-36vJ_0+1728\Big(2J_1-J_0^2-\frac29G\Big)v^2\Big],\nonumber\\
    \beta _z&=&-(2\sigma-D)z \Big[1-24vJ_0-2zJ_0\nonumber\\
    &+&576(2J_1-J_0^2-\frac 23G)v^2\Big],\nonumber\\
    \beta _w&=&-(2\sigma-D)w \Big[1-24vJ_0-4zJ_0+2wJ_0\nonumber\\
    &+&576(2J_1-J_0^2-\frac 23G)v^2\Big].\nonumber\\
    \gamma_t&=&(2\sigma-D)\Big[-12vJ_0-2zJ_0+2wJ_0\nonumber\\
    &+&288\Big(2J_1-J_0^2-\frac13G\Big)v^2\Big], \nonumber\\
    \gamma_\varphi&=&(2\sigma-D)192Gv^2,\nonumber\\
    J_1&=&\int \frac{d^Dqd^Dp}{(1+|\vec{q}|^\sigma)^2(1+|\vec{p}|^\sigma)(1+|q^2+p^2+2\vec{p}\vec{q}|^{\sigma/2})},\nonumber\\
    J_0&=&\int \frac{d^Dq}{(1+|\vec{q}|^\sigma)^2},\nonumber\\
    G&=&-\frac{\partial}{\partial |\vec{k}|^\sigma}\int d^Dq
    d^Dp(1+|q^2+k^2+2\vec{k}\vec{q}|^\sigma)^{-1}\nonumber\\
    &&(1+|\vec{p}|^\sigma)^{-1}(1+|q^2+p^2+2\vec{p}\vec{q}|^{\sigma/2})^{-1}\nonumber
\end{eqnarray}
Redefining the effective interaction vertices as
\begin{equation}\label{vertex}
    v_1=v\cdot J_0,\ \ \ \ \
    v_2=z\cdot J_0,\ \ \ \ \  v_3=w\cdot J_0.
\end{equation}
we arrive at the following expressions for the $\beta-$ and $\gamma$-
functions:
\begin{eqnarray}\label{beta}
    \beta_1&=&-(2\sigma-D)v_1\Big[1-36v_1\\
    &&+1728\Big(2\widetilde{J_1}-1-\frac29\widetilde{G}\Big)v_1^2\Big],\nonumber\\
    \beta _2&=&-(2\sigma-D)v_2\Big[1-24v_1-2v_2\nonumber\\
    &&+576(2\widetilde{J_1}-1-\frac 23\widetilde{G})v_1^2\Big],\nonumber\\
    \beta _3&=&-(2\sigma-D)v_3\Big[1-24v_1-4v_2+2v_3\nonumber\\
    &&+576(2\widetilde{J_1}-1-\frac 23\widetilde{G})v_1^2\Big],\nonumber\\
    \gamma_t&=&(2\sigma-D)\Big[-12v_1-2v_2+2v_3\nonumber\\
    &&+288\Big(2\widetilde{J_1}-1-\frac13\widetilde{G}\Big)v_1^2\Big], \nonumber\\
    \gamma_\varphi&=&(2\sigma-D)192\widetilde{G}v_1^2.\nonumber
\end{eqnarray}

The redefining procedure makes sense for a $\sigma\leq D/2$. In
this case, $J_0$, $J_1$, and $G$ become divergent functions.
Introducing the cutoff parameter $\Lambda$ and considering the
ratios $J_1/J_0^2$, $G/J_0^2$ in the limit of
$\Lambda\rightarrow\infty$, we obtain finite expressions.

The values of the integrals were determined numerically.
For the case a $a\leq D/2$, a sequence of the values of
$J_1/J_0^2$ and $G/J_0^2$ was constructed for different $\Lambda$ and
then approximated to infinity.

It is well known that perturbative series expansions are
asymptotic and the interaction vertices of the order parameter
fluctuations in the fluctuation region are suf- ficiently large to
directly apply Eqs. (7). Therefore, to extract the necessary
physical information from the expressions derived above, the
Pade–Borel method generalized to the three-parameter case was
used. The corresponding direct and inverse Borel transformations
have the form
\begin{equation}
\begin{array}{rl} \displaystyle
  & f(v_1,v_2,v_3)=\sum\limits_{i_1,i_2,i_3}c_{i_1,i_2,i_3}v_1^{i_1}v_2^{i_2}v_3^{i_3}\nonumber\\
  &=\int\limits_{0}^{\infty}e^{-t}F(v_1t,v_2t,v_3t,v_4t)dt,  \\
  & F(v_1,v_2,v_3)=\sum\limits_{i_1,i_2,i_3}\frac{\displaystyle c_{i_1,i_2,i_3}}{\displaystyle(i_1+i_2+i_3)!}v_1^{i_1}v_2^{i_2}v_3^{i_3}.
\end{array}
\end{equation}

For an analytical continuation of the Borel transform of
a function, a series in an auxiliary variable è is introduced:
\begin{equation}  \displaystyle
   {\tilde{F}}(v_1,v_2,v_3,\theta)=\sum\limits_{k=0}^{\infty}\theta^k\sum\limits_{i_1,i_2,i_3}\frac{\displaystyle c_{i_1,i_2,i_3}}{\displaystyle k!}v_1^{i_1}v_2^{i_2}v_3^{i_3}\delta_{i_1+i_2+i_3,k}\  ,
\end{equation}
and the $[L/M]$ Pade approximation is applied to this
series at the point $\theta=1$. This approach was proposed
and tested in [14] in describing the critical behavior of
systems characterized by several vertices of interaction
of the order parameter fluctuations. The property [14]
of the system retaining its symmetry under the Pade
approximants in the variable è is essential in the
description of multivertex models.

In the two-loop approximation, the â functions were
calculated using the [2/1] approximant. The character
of the critical behavior is determined by the existence
of a stable fixed point satisfying the set of equations
\begin{equation}\label{nep}
    \beta_i(v_1^*,v_2^*, v_3^*)=0 \ \ \ \ (i=1,2,3).
\end{equation}
The requirement that the fixed point be stable is
reduced to the condition that the eigenvalues $b_i$ of the
matrix
\begin{equation}  \displaystyle
B_{i,j}=\frac{\partial\beta_i(v_1^*,v_2^*, v_3^*)}{\partial{v_j}}.
\end{equation}
lie in the half-plane of the right-hand complex. The
fixed point with $v^* = 0$, which corresponds to the tricritical
behavior, is a saddle point and must be stable in
the directions determined by the variables $z$ and $w$ and
unstable in the direction determined by the variable $v$.
The stabilization of the tricritical fixed point in the
direction determined by the variable $v$ is achieved by
taking into account the sixth-order terms with respect to
the order parameter fluctuations in the effective Hamiltonian
of the model. The fixed point with $z^* = w^*$,
which corresponds to the tricritical behavior of the second
type, is also a saddle point and must be stable in the
directions determined by the variables $v$ and $z$ and
unstable in the direction determined by the variable $w$.
Its stabilization is possible at the expense of the anharmonic
effects.

The resulting set of the resummed $\beta$ functions contains
a wide variety of fixed points. The table specifies
the fixed points that are of most interest for describing
the critical and tricritical behavior and lie in the physical
region of vertex values with $v$, $z$, $w\geq0$. The table
also shows the eigenvalues of the stability matrix for
the corresponding fixed points and the critical indices $\nu$
and $\eta$.

The analysis of the values and stability of the critical
points suggests the following conclusions. Qualitatively,
the critical phenomena seem to be identical for
any value of the long-range interaction parameter $a$.
The critical behavior of incompressible systems is
unstable with respect to the deformation degrees of
freedom (points 1). The stable point proves to be the
one at a constant strain (points 2). Fixed points 3
describe the first type of tricritical behavior of compressible
systems, which occurs at constant pressure.
Fixed points 4 are tricritical for systems studied at constant
volume. Points 5 are fourth-order critical points at
which two tricritical curves intersect.

For the tricritical behavior of the first type (points 3),
Hamiltonian (5) is isomorphic with the Hamiltonian of
an incompressible homogeneous model and, hence, the
critical indices also coincide with those of the incompressible
model. The tricritical behavior of the second
type (points 4) corresponds to the critical behavior of a
spherical model and is determined by the corresponding
indices. The fourth-order fixed points (points 5) are
characterized by the field-average values of the critical
indices.

The large values of the effective vertices $z$ and $w$ in
comparison with the systems with short-range interactions
[3] are caused by the fact that the mechanism governing
the effect of elastic deformations on the critical
phenomena is related to the dependence of the interaction
integral in the Ising model on the distance between
the lattice sites.

The study described above revealed the considerable
effect of elastic deformations on the critical behavior
of systems with a long-range interaction. This effect
manifests itself as a change in the values of the critical
indices for Ising systems along with the appearance of
multicritical points in the phase diagrams of the substances.

The work is supported by Russian Foundation for Basic Research N 04-02-16002.

\begin{table*}
\begin{center}
\begin{tabular}{|c|c|c|c|c|c|c|c|c|} \hline
 N & $v_1^{*}$  &$ v_2^{*}$   &$v_3^{*}$    &$b_{1}$    &$b_{2}$    &$b_{3}$     &$\nu$      & $\eta$   \\
\hline
\multicolumn{9}{|c|} {$\sigma=1.9$} \\
\hline
 1 & 0.042067 &  0        &  0        &  0.684    & -0.184    & -0.184     & 0.652688  & 0.113420 \\
 2 & 0.044353 &  0.095190 &  0        &  0.684    &  0.185    &  0.183     & 0.751433  & 0.113420 \\
 3 & 0.044353 &  0.095190 &  0.095190 &  0.684    &  0.185    & -0.185     & 0.652688  & 0.113420 \\
 4 & 0        &  0.5      &  0        &     -1    &  1        &  1         & 1.052632  &  0.1       \\
 5 & 0        &  0.5      &  0.5      &     -1    &  1        & -1         & 0.526316  &  0.1       \\
\hline
\multicolumn{9}{|c|} {$\sigma=1.8$} \\
\hline
 1 & 0.023230 &  0        &  0        &  0.628    & -0.488    & -0.488     & 0.636349  & 0.207461 \\
 2 & 0.023230 &  0.245404 &  0        &  0.628    &  0.489    &  0.490     & 0.782912  & 0.207461 \\
 3 & 0.023230 &  0.245404 &  0.245404 &  0.628    &  0.489    & -0.489     & 0.636349  & 0.207461 \\
 4 & 0        &  0.5      &  0        &     -1    &  1        &  1         & 1.111111  &  0.2       \\
 5 & 0        &  0.5      &  0.5      &     -1    &  1        & -1         & 0.555556  &  0.2       \\\hline
\multicolumn{9}{|c|} {$\sigma=1.7$} \\
\hline
 1 & 0.020485 &  0        &  0        &  0.699    & -0.532    & -0.532     & 0.667452  & 0.304862 \\
 2 & 0.020485 &  0.266497 &  0        &  0.699    &  0.533    &  0.532     & 0.830648  & 0.304862 \\
 3 & 0.020485 &  0.266497 &  0.266497 &  0.699    &  0.533    & -0.533     & 0.667452  & 0.304862 \\
 4 & 0        &  0.5      &  0        &     -1    &  1        &  1         & 1.176471  &  0.3       \\
 5 & 0        &  0.5      &  0.5      &     -1    &  1        & -1         &  0.588235 &  0.3       \\\hline
\multicolumn{9}{|c|} {$\sigma=1.6$} \\
\hline
 1 & 0.015974 &  0        &  0        &  0.874    & -0.616    & -0.616     & 0.697361  & 0.403936 \\
 2 & 0.015974 &  0.309684 &  0        &  0.874    &  0.617    &  0.620     & 0.889473  & 0.403936 \\
 3 & 0.015974 &  0.309684 &  0.309684 &  0.874    &  0.617    & -0.618     & 0.697361  & 0.403936 \\
 4 & 0        &  0.5      &  0        &     -1    &  1        &  1         &  1.25     &  0.4       \\
 5 & 0        &  0.5      &  0.5      &     -1    &  1        & -1         &  0.625    &  0.4       \\\hline
\end{tabular} \end{center} \end{table*}
\newpage
\def\baselinestretch{1.0}

\end{document}